\documentclass[12pt]{article}
 \input epsf.sty

\usepackage{latexsym}
\usepackage{graphics}

\usepackage {amsfonts}

\begin{document}

\def\ket{\rangle}
\def\bra{\langle}
\def\CA{{\cal A}}
\def\CB{{\cal B}}
\def\CC{{\cal C}}
\def\CD{{\cal D}}
\def\CE{{\cal E}}
\def\CF{{\cal F}}
\def\CG{{\cal G}}
\def\CH{{\cal H}}
\def\CI{{\cal I}}
\def\CJ{{\cal J}}
\def\CK{{\cal K}}
\def\CL{{\cal L}}
\def\CM{{\cal M}}
\def\CN{{\cal N}}
\def\CO{{\cal O}}
\def\CP{{\cal P}}
\def\CQ{{\cal Q}}
\def\CR{{\cal R}}
\def\CS{{\cal S}}
\def\CT{{\cal T}}
\def\CU{{\cal U}}
\def\CV{{\cal V}}
\def\CW{{\cal W}}
\def\CX{{\cal X}}
\def\CY{{\cal Y}}
\def\CZ{{\cal Z}}

\newcommand{\todo}[1]{{\em \small {#1}}\marginpar{$\Longleftarrow$}}
\newcommand{\labell}[1]{\label{#1}}
\newcommand{\bbibitem}[1]{\bibitem{#1}\marginpar{#1}}
\newcommand{\llabel}[1]{\label{#1}\marginpar{#1}}
\newcommand{\dslash}[0]{\slash{\hspace{-0.23cm}}\partial}

\newcommand\Z{{\mathbb Z}}   
\newcommand\R{{\mathbb R}}   
\newcommand\C{{\mathbb C}}   
\renewcommand\H{{\mathbb H}}   
\newcommand\K{{\mathbb K}} 
\renewcommand\O{{\mathbb O}}   
\renewcommand\S{{\mathbb S}}      
\newcommand\RP{{\mathbb {RP}}}   
\newcommand\HP{{\mathbb {HP}}}   
\newcommand\KP{{\mathbb {KP}}}   
\newcommand\OP{{\mathbb {OP}}}   
\renewcommand\P{{\mathbb P}}
\newcommand\B{{\mathbb B}}

\newcommand{\sphere}[0]{{\rm S}^3}
\newcommand{\su}[0]{{\rm SU(2)}}
\newcommand{\so}[0]{{\rm SO(4)}}
\newcommand{\bK}[0]{{\bf K}}
\newcommand{\bL}[0]{{\bf L}}
\newcommand{\bR}[0]{{\bf R}}
\newcommand{\tK}[0]{\tilde{K}}
\newcommand{\tL}[0]{\bar{L}}
\newcommand{\tR}[0]{\tilde{R}}

\newcommand{\btzm}[0]{BTZ$_{\rm M}$}
\newcommand{\ads}[1]{{\rm AdS}_{#1}}
\newcommand{\ds}[1]{{\rm dS}_{#1}}
\newcommand{\eds}[1]{{\rm EdS}_{#1}}
\newcommand{\sph}[1]{{\rm S}^{#1}}
\newcommand{\gn}[0]{G_N}
\newcommand{\SL}[0]{{\rm SL}(2,R)}
\newcommand{\cosm}[0]{R}
\newcommand{\hdim}[0]{\bar{h}}
\newcommand{\bw}[0]{\bar{w}}
\newcommand{\bz}[0]{\bar{z}}
\newcommand{\pat}{\partial}
\newcommand{\lp}{\lambda_+}
\newcommand{\bx}{ {\bf x}}
\newcommand{\bk}{{\bf k}}
\newcommand{\bb}{{\bf b}}
\newcommand{\BB}{{\bf B}}
\newcommand{\tp}{\tilde{\phi}}
\hyphenation{Min-kow-ski}

\def\apr{\alpha'}
\def\str{{str}}
\def\lstr{\ell_\str}
\def\gstr{g_\str}
\def\Mstr{M_\str}
\def\lpl{\ell_{pl}}
\def\Mpl{M_{pl}}
\def\varep{\varepsilon}
\def\del{\nabla}
\def\grad{\nabla}
\def\tr{\hbox{Tr}}
\def\perp{\bot}
\def\half{\frac{1}{2}}
\def\p{\partial}
\def\perp{\bot}
\def\eps{\epsilon}
\newcommand{\be}{\begin{equation}}
\newcommand{\ee}{\end{equation}}
\newcommand{\ben}{\begin{eqnarray}}
\newcommand{\een}{\end{eqnarray}}
\def\dalemb#1#2{{\vbox{\hrule height .#2pt
        \hbox{\vrule width.#2pt height#1pt \kern#1pt
                \vrule width.#2pt}
        \hrule height.#2pt}}}
\def\square{\mathord{\dalemb{6.8}{7}\hbox{\hskip1pt}}}

\newcommand{\bea}{\begin{eqnarray}}
\newcommand{\eea}{\end{eqnarray}}

\renewcommand{\thepage}{\arabic{page}}
\setcounter{page}{1}

\rightline{hep-th/0206180}
\rightline{UPR-1001-T,NI02017-MTH}
\vskip 1cm
\centerline{\Large \bf Brane Resolution
and Gravitational Chern-Simons terms }
\vskip 0.5 cm
\renewcommand{\thefootnote}{\fnsymbol{footnote}}
\centerline{{\bf Francisco A. Brito$^1$,\footnote{fabrito@hep.upenn.edu}
Mirjam Cveti\v c$^{1,2}$,\footnote{cvetic@cvetic.hep.upenn.edu}
and
Asad Naqvi$^1$\footnote{naqvi@rutabaga.hep.upenn.edu}
}}
\vskip .5cm
\centerline{\it $^1$David Rittenhouse Laboratories, University of
Pennsylvania}
\centerline{\it Philadelphia, PA 19104, U.S.A.}
\vskip .2cm
\centerline{\it $^2$ Isaac Newton Institute for Mathematical Sciences,
University of Cambridge,}
\centerline{Cambridge, CB3 0EH, U.K.}

\setcounter{footnote}{0}
\renewcommand{\thefootnote}{\arabic{footnote}}

%



\begin{abstract}
We show that  gravitational Chern-Simons corrections,
associated with the sigma-model anomaly on the M5-brane
world-volume,
 can    resolve the
M2-brane solution with Ricci-flat, special holonomy 
transverse space.  
We explicitly find smooth solutions in the cases 
when the transverse space is a manifold
of Spin(7) holonomy  and SU(4) holonomy. We comment on the consequences of
these
results
for the holographically related three-dimensional theories living on the world volume of a 
stack of such resolved  M2-branes.

\end{abstract}



\section{Introduction}
The original AdS/CFT correspondence is an equivalence between
Type IIB strings moving in the near horizon geometry created by
a stack of $N$ D3 branes and  $\CN=4$ SU(N) gauge theory which is the 
infra-red limit of the theory living on the world volume of the branes
\cite{adscft}. In the large $N$ limit, with $g_{\rm YM}^2 N$ fixed but large, 
tree level supergravity is a good approximation to string theory. 
The $\apr$ and $g_s$ expansions in string theory map
to ${1 /\sqrt{g_{\rm YM}^2N}}$ and ${1/N}$ expansions in
the gauge theory.  
Stringy effects in AdS spaces are hard to compute since
it is difficult to  quantize strings
 in RR backgrounds. However, we know some $\apr$ 
corrections to the low energy space-time effective action in the form
of higher curvature corrections of the form $\apr^3 R^4$.
Such an 
$\apr$ expansion should contain information about the strong coupling
expansion of the gauge theory \cite{tseytlinfrolov}.

We can study similar questions in the context of M-theory 
on $\ads{4} \times X^{7}$ with AdS scale $R \approx  l_p N^{1/6}$. In 
this setup, $l_p/R$ corrections to supergravity go like $1/N^{1/6}$. 
In this paper, we consider higher curvature corrections to 11 dimensional
supergravity action which are the gravitational Chern-Simons 
terms arising from the $\sigma$-model anomaly on the M5
brane world volume, thereby modifying the Bianchi identity 
for the five brane. Such corrections, schematically go like $l_p^6 R^4$
and hence correspond to $1/N$ effects in the corresponding holographic
field theories. Specifically, we consider the correction to
classical M2-brane solutions arising from such gravitational 
Chern-Simons terms, when such M2-branes have transverse
spaces which are Ricci-flat deformations of cones over seven
dimensional Einstein manifolds. Branes at such conical
singularities have been studied extensively for the past
few years and we briefly recap some of the features salient for our study. 

In  the search of a string dual for four dimensional
gauge theories with lesser supersymmetry, 
variations of the original correspondence involve placing D3 branes
at the apex of a Ricci flat six dimensional cone whose base is a five
dimensional Einstein manifold $X^5$ \cite{conifold}. This leads
to the conjecture that type IIB string theory on
$\ads{5} \times X^5$ is dual to the low energy limit of the
theory on the D3 branes at the singularity.
For the conifold singularity, $X^5$ is the Einstein space
$T^{1,1}$. 
 Adding
fractional D3 branes, (D5 branes wrapped over two-cycles of $X_5$)
introduces a non-zero three-form RR flux
through the three-cycle of $T^{1,1}$ and results in a non-AdS bulk
solution.  Correspondingly, on the gauge theory side, fractional
branes
lead to a non-conformal gauge theory with running couplings. 
Supergravity solutions for such configurations of
D3 branes were considered in \cite{kn,kt}. The solution in \cite{kt} that includes
fractional D3-brane,  was singular.  In \cite{ks}, the singularity was
resolved
by replacing the singular conifold by  a (smooth) deformed conifold
\cite{candelas}. This regular
supergravity solution  
realized the chiral symmetry breaking and confinement of the
 dual $\CN=1$ supersymmetric four-dimensional gauge theory geometrically (see
also \cite{carlos}. 

It is possible to generalize the above setup to other p-branes
placed at tips of other cones, {\em i.e.} the transverse space
to the p-branes is a cone $ds^2=dr^2+r^2ds_X^2$, 
where $ds_X^2$ is the metric on an Einstein manifold $X$, 
which is called the base of the cone \cite{conical}. Turning
on additional $F_{6-p}$ field strengths corresponds to
deforming the holographically related field theory.  
In \cite{clp,cglp} (for a review see \cite{rev}), 
it was shown that after addition of such fluxes 
 regular solutions can be obtained which have the feature that
the singular conical transverse space is resolved in the IR
region  of the field theory.  Also, these solutions generically do 
not have horizons, implying that  a mass gap has been generated in
the dual field theory. 

In finding smooth, non-singular solutions, a
crucial role
is played by the Chern-Simons type terms with additional
field strengths   which modify the Bianchi identities and/or equations
of motions for the original field strength. Note that the additional field
strengths $F_{6-p}$ are supported by harmonic forms on the special
holonomy space, and in particular the $L^2$ normalizability of these
harmonic  forms ensures that the solutions  are regular both in the
interior and at large radial coordinate $r$. For most of the cases and in
particular for
resolved M2-branes and D2 branes  \cite{clp,cglp,cglp3,cglp3p}, 
with the transverse space
asymptotically  conical (AC)  Spin(7) and with $G_2$ holonomy,
the asymptotic form of the field strength was such that
it did not produce any new flux at infinity. As such, 
these configurations do not describe gravity dual of 
fractional branes \cite{kh,cglp3p} but perturbations of the boundary
field theory living on a large stack of regular branes by relevant operators. 
The power law fall-off of the field strength determines the dimension of the
operator to be added to the gauge theory action.

In this paper we focus on another aspect of resolved brane solutions 
which is  due to the  gravitational
Chern-Simons type corrections. These terms, first 
established within Type II string
theory \cite{witten},  have  their M-theory analog due to 
$\sigma$-model anomaly on M5-brane. 
Within eleven dimensional supergravity we shall find explicit solutions
for resolved M2-brane due to these  higher derivative corrections. 
\footnote{There are of course other higher derivative corrections
which we are not considering in this paper. 
However, since the
gravitational Chern Simons term is special in the sense that it corresponds to a bulk charge being induced\cite{hawking}. We believe that the other higher derivative corrections will not change the qualitative behavior we describe in this paper.  }

Some of the aspects  in connection with  M2-branes  and gravitational
Chern-Simons terms were studied earlier in \cite{duff,beckers,hawking,becker2,gukov2}.
We extend this analysis by finding explicit solutions for
such M2-branes whose  eight-dimensional transverse
space  is a   Ricci-flat deformation of cones over seven
dimensional Einstein manifolds. 
These are non-compact, smooth spaces with special Spin(7) and SU(4)
holonomy, i.e.
Ricci flat spaces with at least one covariantly constant spinor, and whose
metrics are explicitly known.
%

The starting point is  the Ansatz  for the original M2-brane solution:
\begin{eqnarray*}
ds^2_{11}&=&H^{-2/3}\eta_{\mu\nu}~dx^\mu dx^\nu+H^{1/3}ds^2_8,\nonumber\\
F_{0123r}&=&\partial_r{H^{-1}}, 
\end{eqnarray*}
where $ds_8^2$ is a Ricci-flat transverse metric of Spin(7) or SU(4)
holonomy. Without
taking into account higher curvature corrections, the function $H$ is 
harmonic, satisfying 
the equation $\Box H=0$ where $\Box$ is the Laplacian on the Ricci-flat
transverse space. $H$ turns out to be singular at the inner boundary of 
the transverse space.  We shall however see that the inclusion of the
gravitational Chern-Simons-type corrections,  which are of the
type $\propto \tr (R^4)
-{1 \over 4} \tr(R^2)^2$, the singularity of the solution can be resolved.
No inclusion of the four-form $G_4$
supported by the special holonomy transverse space is needed. 
In addition, the gravitational Chern-Simons term induces a bulk M2 
brane charge and the solution asymptotically approaches $\ads{4} \times
X^7$ where $X^7$ is the base of the transverse cone. 
However, since the only scale in the problem is $l_p$,  the solution has a characteristic
curvature scale of order $l_p$ and thus supergravity approximation cannot
be trusted. As we will see later, to get a good supergravity description, we indeed
need to turn on four-form field strength $G_4$ in such a way that the curvature 
of the solution is everywhere much smaller than $l_p$. 

The paper is organized as follows. In section \ref{Chern}, we discuss
generalities about the gravitational Chern-Simons eight-form $\sigma$
model anomaly.
 In
section \ref{Spin} ,taking into account the correction
to the M2-brane equations of motion from this term, we explicitly construct 
smooth M2-brane solutions for the case with the
transverse space is one of 
 two different metrics of Spin(7) holonomy, one originally
constructed in \cite{bs,gpp,pp} and the other recently found in 
\cite{cglp2,cglpm,cglp4}
which is
asymptotically locally conical (ALC). The  corresponding dual 
(2+1)-dimensional  field theories have $\CN=1$ supersymmetry.
 In section  \ref{Stenzel}, we repeat the analysis when
the transverse space is  $T^*S^4$, with  Stenzel metric
which has SU(4) holonomy. The holographically dual (2+1)-dimensional field
theory on the world volume of the M2-brane has $\CN=2$ supersymmetry.
In concluding section \ref{Conclusions}, we comment on the
interpretation
of our result in the dual field theory. 
In  Appendix A we present the details of the  calculation for the Ricci
tensor and curvature for a class of Spin(7) holonomy metric and in Appendix B
we collected the details for the calculation of the harmonic functions.
 
\section{Gravitational
Chern-Simons
corrections and M2-branes} \label{Chern}
 
The bosonic sector of $d=11$ supergravity \cite{cjs} is given by 
\ben
\label{action}
S_{11}={1 \over 2} \int{d^{11}x\sqrt{g}R}-\int{\left(\frac{1}{2}F\wedge *F
+\frac{1}{6}A\wedge F\wedge F\right)},
\een
where $g_{MN}$ is the space time metric and $A$ is a
three form with field strength
$F=dA$. The field strength obeys the Bianchi identity
$dF=0$ and its equation of motion is 
\[
d*F=-{1 \over 2} F \wedge F. 
\]
The gravitational Chern-Simons corrections associated with
the $\sigma$-model anomaly on the M5-brane \cite{duff,witten}, modify the
equation:
\ben
\label{Seq01}
d{*}F=-\frac{1}{2}F\wedge F+(2\pi)^4\beta\, {\CX_8},
\een
where $\beta$ is related to the five-brane tension as 
$T_6\!=\!\beta/(2\pi)^3$
and the eight-form anomaly ${\CX_8}$ can be expressed in terms of
the
curvature two-form $\Theta$:
\ben
\label{Seq02}
{ \CX_8}=\frac{1}{(2\pi)^4}\left\{-\frac{1}{768}({\rm Tr}\,\Theta^2)^2+
\frac{1}{192}({\rm Tr}\,\Theta^4)\right\}.
\een
The equation of motion (\ref{Seq02}) can be derived from the action
(\ref{action}) with the addition of the following term: 
\ben
\Delta S_{11}=\int{A\wedge\left\{-\frac{1}{768}({\rm Tr}\,\Theta^2)^2+
\frac{1}{192}({\rm Tr}\,\Theta^4)\right\}}.
\een

We look for solutions with (2+1)-dimensional Lorentz invariance:
\ben
\label{Ansatz}
ds^2_{11}&=&H^{-2/3}dx^\mu dx^\nu\eta_{\mu\nu}+H^{1/3}ds^2_8,\nonumber\\
F&=&d^3x\wedge dH^{-1}+mG_4.
\een
Here,  we have also added a harmonic four-form $G_4$ on the transverse
eight-manifold. 
This ansatz preserves supersymmetry as shown in \cite{hawking} if
\ben
\label{KG}
{\square\, H}=-\frac{1}{48}m^2|G_{(4)}|^2+(2\pi)^4\beta\, X_8,
\een
where $\CX_8=X_8{\rm dvol}_8$,
${\rm dvol}_8$
is the volume form and $\Box$ is the Laplacian on the eight-fold.
Such solutions were studied in 
\cite{clp,cglp}. For vacuum solutions $m=0$ and if $X_8$ is trivial, 
 we can choose $H$ to be a harmonic function on the eight-fold.  In fact, a
constant $H$ 
will then be a solution, whence we get a product manifold with metric 
$ds^2=dx^\mu dx^\nu \eta_{\mu \nu}+ds_8^2$. 
However, if the holonomy of the eight-manifold is non-trivial, the anomaly 
term may not vanish, and $H={\rm constant}$ is not a solution anymore. 
This can be interpreted as a distribution of background charge over the
eight-fold, induced by the anomaly term.

As we will see, a smooth solution
for $H$ can 
be found even when the four-form $G_4$ {\it is not} turned on. 
However,
such 
solutions fail to have a good AdS/CFT interpretation since the
solution
has a curvature scale of order $l_p$ and  supergravity is not
a good description. To get a good gravity description, we can 
turn 
on background fluxes through $G_4$ which is an (anti) self-dual
four-form, supported on the special holonomy space, such that the length
scale associated with the metric is 
everywhere $\gg l_p$.  

In what follows, we will
 study specific examples of transverse
eight manifolds to be 
(i)  two different manifolds of Spin(7) holonomy (one is the original one
\cite{bs,gpp}
 with AC structure and another set, recently
constructed in \cite{cglp2,cglpm}, has ALC structure)  
where the (2+1)-dimensional field theory has $\CN=1$ SUSY,
(ii) $T^*S^4$, with Stenzel metric,  which has
SU(4) holonomy, and the (2+1)-dimensional field theory on the world
volume
of the M2-brane has $\CN=2 $ SUSY.

\section{Resolved M2-brane and Spin(7) holonomy}\label{Spin}
In this section, we will find explicit M2-brane solutions when the
transverse space is a manifold of Spin(7) holonomy. As such, 
there is one covariantly constant spinor on the manifold and 
the corresponding holographic (2+1)-dimensional field theory
has $\CN=1 $ supersymmetry. We consider the general
Ansatz for a Spin(7) manifold introduced in \cite{cglp2}, 
with special cases yielding metrics constructed in \cite{bs,gpp}: 
\ben
\label{newmet}
ds_8^2=h^2\,dr^2+a^2({D\mu^i})^2+b^2\sigma^2+c^2d\Omega_4^2,\qquad i=1,2,3
\een
where $h,\,a,\,b$ and $c$ are functions of a radial coordinate $r$,  $\mu_i$
parameterize an $\sph{2}$ and satisfy $\mu^i\mu^i\!=\!1$,
\ben
\mu_1=\sin{\theta}\sin{\psi},\qquad \mu_2=\sin{\theta}\cos{\psi},\qquad 
\mu_3=\cos{\theta}.
\een
and
\ben
D\mu_i\!=\!d\mu_i+\epsilon_{ijk}A^j\,u^k,\qquad\sigma=d\varphi
+{\cal A}_{},\qquad
{\cal A}_{}\equiv \cos{\theta}d\psi-\mu^iA_{}^i.
\een
The 1-form $A_{}^i$ is the $SU(2)$ Yang-Mills instanton on $S^4$.
In terms of coordinates $(\theta , \psi)$ on $S^2$, we have
\begin{eqnarray}
\sum_i (D\mu^i)^2&\!=\!&(d\theta-A_{}^1\cos\psi+A_{}^2 \sin\psi)^2 \nonumber \\
& & \,\,\,\,\,\,\,\,+\sin^2\theta
( d\psi+A_{}^1 \cot\theta \sin \psi + A_{}^2 \cot\theta \cos \psi-A_{}^3)^2.
\end{eqnarray}
The Vielbeine are given by
\begin{eqnarray}
\hat{e}^{0}&\!=\!&h\,dr,\,\,\,\,\,\,\, \qquad\hat{e}^\alpha=c\,e^{\alpha},\nonumber \\
\qquad\hat{e}^1&\!=\!&a(d\theta-A_{}^1\cos\psi+A_{}^2 \sin\psi),\nonumber \\
\qquad\hat{e}^2&\!=\!& a \sin \theta( d\psi+A_{}^1 \cot\theta \sin \psi + A_{}^2 
\cot\theta \cos\psi-A_{}^3),
\nonumber \\
\qquad\hat{e}^3 &\!=\!& b \,\sigma, \label{Sp7vp}
\end{eqnarray}
where $e^\alpha$, with $\alpha\!=\!4,5,6,7$, is an orthonormal
basis of the tangent-space 1-forms on the unit $S^4$.

The spin connection $\omega_{ab}$ satisfying $de^a+\omega^a{}_b \wedge e^b=0$
and the curvature two form $\Theta_{ab}=d\omega_{ab}+\omega_{a}{}^c\wedge
\omega_{cb}$ for this are given in Appendix A.  
In what follows, we will study two cases of such manifolds with Spin(7)
holonomy. 
\subsection{Old Spin(7) holonomy space}
For the special case when $a=b$, the Ansatz given in (\ref{newmet})
reduces to \cite{bs,gpp,pp}:
\be
ds_8^2=h^2 dr^2+a^2(\sigma_i -A^i_{})^2+c^2d \Omega_4^2,
\label{oldmet}\ee
where we have used the relation
\be
\sum_i(\sigma_i -A^i_{})^2=\sum_i (D\mu^i)^2+\sigma^2.
\ee
Conditions for Ricci flatness and Spin(7) holonomy for this Ansatz have the following
solution\cite{bs,gpp,pp}:
\ben
\label{old}
h^2(r)=\Big(1-\frac{{\it l}^{10/3}}{r^{10/3}}\Big)^{-1},\;\; a^2(r)=b^2(r)=
\frac{9}{100}r^2\Big(1-\frac{{\it l}^{10/3}}{r^{10/3}}\Big),\;\; c^2(r)=
\frac{9}{20}r^2.
\een
This metric is a resolution of a cone with a squashed seven sphere
base. (Indeed, when $l=0$, (\ref{old}) becomes $ds_8^2=dr^2+r^2
ds^2_{{\rm squashed} ~\sph{7}}$.) 
The  space is asymptotically conical (AC)  with the  principal
orbits $S^7$, viewed as an $S^3$ bundle over $S^4$.

\paragraph{Singular cone (${l=0}$):}
By placing M2-branes at the tip
of this cone and taking an appropriate scaling limit, we can 
arrive at a correspondence between M-theory on $\ads{4}
\times S^7_{\rm Squashed}$ and the $\CN=1$ field theory living on the
world volume of such M2-branes. The scaling limit corresponds
to looking for the solution to the equation of motion $\Box H=c\delta(r)$
which approaches zero asymptotically. Notice that there is a $\delta$
function source term on the right hand side of the EOM. The solution 
thus obtained is $H={32 \pi^2 N l_p^6  \over r^6}$ for $N$ M2-branes
placed at the singularity. The space-time metric is $\ads{4} \times S^7_{\rm squashed}$, 
with the AdS scale given by $R=(32 \pi^2 N)^{1/6}l_p$.  
\vspace{-0.2cm}
\paragraph{Resolved cone with Spin(7) holonomy  (${ l \neq 0}$):}
We will now find M2-branes solutions with transverse space 
 a smooth Spin(7) holonomy
manifold given by (\ref{old}), arising from resolution of the conical
singularity by replacing the singular "tip" of the cone
by a bolt. As we will discuss later, such smooth solutions are
gravity dual of $\CN=1$ field theory living on the world volume 
of the M2-branes perturbed with relevant operators (associated with the
pseudoscalar fields of the dual field theory \cite{kh}).

We will first look for vacuum solutions, {\em i.e.} solutions with
no  four-form flux turned on ($m=0$). There is, 
however, a four-form bulk charge induced by the anomaly term 
$X_8$ which is now non-zero:
\ben
\label{x8s7}
&&(2\pi)^4\,X_8=\frac{20\,l^3}{3^7}(1530\,r^{20/3}\,l^{31/3}+3120\,r^{10}\,l^7
-1228\,r^{50/3}\,l^{1/3}-697\,l^{17}\nonumber \\ & & \,\,
-1185\,r^{40/3}\,l^{11/3}-1540\,l^{41/3}\,
r^{10/3})\Big/r^{64/3}(-r^{10/3}+l^{10/3})^2,
\een
The equation of motion (\ref{KG}) can be solved explicitly and details are 
given in Appendix B.
The solution is in general is singular at $r=l$. However, we can remove
this singularity and find a
smooth solution if we choose a specific  integration constant  as discussed
in Appendix B.
The full regular solution is
\ben
\label{Seq5.21p}
H(r)&=&\frac{\beta}{34904520}(-5758444\,l^4\,r^{32/3}+23942926\,r^{14}\,l^{2/3}-11848824\,l^{22/3}\,r^{22/3}\nonumber\\
& &-834309\,l^{14}\,r^{2/3}-5501349\,l^{32/3}\,r^4)\Big/l^{2/3}
\,r^{50/3}\,(r^{10/3}-l^{10/3})+c_2. 
\een
Asymptotically, as $r\to\infty$  we have
\ben
\label{Seq6.222p}
H(r)\sim \frac{90011}{131220}\frac{\beta}{r^6}+\frac{53171}{102060}
\frac{\beta\,l^{10/3}}{r^{28/3}}+...+c_2.
\een
As is usual in the AdS/CFT correspondence, we will choose $c_2=0$. 
The space is asymptotically $\ads{4} \times S^7_{\rm squashed}$. The
length scale of AdS is $O(l_p)$ so we cannot trust supergravity. 

We can, however, 
get an AdS radius $\gg l_p$, by turning on 
an anti-self-dual harmonic four-form \cite{clp}:
\ben
|G_{(4)}|^2=\frac{35840000\,l^{4/3}}{729\,r^{28/3}}.
\een
Now the equation of motion (\ref{KG}) for $H$ has a source
term arising from $G_4$ as well as $X_8$. 
Solving for $H$ \cite{clp}, we find a smooth solution which
near $r \approx l$ and $r\rightarrow \infty$ behaves:
\begin{enumerate}
\item
As $r\rightarrow l$ we find
\ben
H(r)\sim c-\left[\frac{8500}{729}\frac{\beta}{l^7}+
\frac{112\times10^3}{729}\frac{m^2}{l^7}\right](r-l)+
\left[\frac{383750}{6561}\frac{\beta}{l^8}+
\frac{952\times10^3}{3^7}\,\frac{m^2}{l^8}\right](r-l)^2\nonumber.
\een
The function $H$ approaches a constant at $r=l$. 
\item
As $r \rightarrow \infty$:
\ben
H(r)=\left[\frac{90011}{131220}\beta+\frac{2\times10^5}{3^7}\,m^2\right]\frac{1}{r^6}-\frac{28\times10^4}{2673}
\frac{l^{4/3}\,m^2}{r^{22/3}}
+\frac{53171}{102060}\frac{\beta\,l^{10/3}}{r^{28/3}}...
\een
\end{enumerate} 
This solution is supposed to describe the gravity dual of the theory living
on the world volume of $N$ M2-branes placed at the conical singularity
perturbed by relevant operators, whose conformal dimension is determined
by the subleading term \cite{kh} in the harmonic function.
Note that the term proportional to
$\beta$  does not contribute at this subleading order. On the other hand 
the leading term  in the harmonic function indeed gives an $\ads{4}
\times S^7_{\rm squashed}$. To get the right AdS scale, we need 
\ben
\frac{90011}{131220}\beta+\frac{2\times10^5}{3^7}\,m^2=32 \pi^2 N l_p^6
\een
The metric has no horizon, implying the existence of 
a mass gap in the dual field theory. 

\subsection{New Spin(7) holonomy space: $\mathbf{\B_8}$}
In \cite{cglp2,cglpm}, new metrics of Spin(7) holonomy, whose structure is
asymptotically locally conical (ALC),  were found by
starting with the  Ansatz (\ref{newmet}), and allowing for the $S^3$
fibers of the old Spin(7) construction   themselves to be ``squashed''.
Namely, the $S^3$ bundle is itself written as a U(1) bundle over $S^2$.
The general two-parameter metrics were given analytically (up to
quadratures)  and analyzed in \cite{cglp2,cglpm}. We
will use one explicit example from these new metrics, namely the
manifold labeled $\B_8$ in \cite{cglp2}. For this manifold,  solution to the Spin(7)
conditions is:
\ben
\label{newc}
&&h^2(r)=\frac{(r-{\it l})^2}{(r-3{\it l})(r+{\it l})},\qquad a^2(r)=\frac{1}{4}(r-3{\it l})
(r+{\it l}), \nonumber\\
&&b^2(r)=\frac{{\it l}^2(r-3{\it l})(r+{\it l})}{(r-{\it l})^2},\qquad
c^2(r)=\frac{1}{2}(r^2-{\it l}^2).
\een
We calculate the spin connection $\omega_{ab}$ and the curvature
two form $\Theta_{ab}$ in Appendix  A.
 The anomaly
eight-form, $\CX_8$ is non-zero. In fact
\ben
\label{newSp}
&&(2\pi)^4\,X_8=-\frac{3\,l^2}{8}(-55\,r^7+491\,r^6\,l-1795\,r^5\,l^2+1871\,r^4\,l^3
+2579\,r^3\,l^4\nonumber\\ & & \,\,
-5431\,r^2\,l^5-1369\,r\,l^6+4733\,l^7)\Big/(r+l)^2\,
(-r+l)^{15}.
\een

In addition, we turn on a self-dual four-form \cite{cglp2}\footnote{In the remainder of this section, we have set $l=1$}:
\ben
\label{newg4}
|G_{(4)}|^2=\frac{96\,(75\,r^6-350\,r^5+829\,r^4-932\,r^3+885\,r^2
-414\,r+99)}
{(r-1)^6\,(r+1)^8}
\een
The explicit solution for the harmonic function  $H$ 
 can be found in Appendix B. Its limits are:
\begin{enumerate}
\item As $r\to 3l$ we have
\ben
H(r)\sim c-\left[\frac{75}{8192}\beta+\frac{9}{2048}\,m^2\right](r-3)+
\left[\frac{141}{8192}\beta+\frac{33}{8192}\,m^2\right](r-3)^2+...
\een
\item As $r \rightarrow \infty$,
\ben
H(r)&=&\left(\frac{10323}{98560}\beta+\frac{63}{20}\,m^2\right)
\frac{1}{r^5}+
\left(\frac{3441}{19712}\beta-\frac{79}{4}\,m^2\right)\frac{1}{r^6}
\nonumber\\
&+&\left(\frac{134199}{137984}\beta
+\frac{317}{4}\,m^2\right)\frac{1}{r^7}
+\left(\frac{34675}{19712}\beta-\frac{953}{4}\,m^2\right)\frac{1}{r^8}+...
\een
\end{enumerate}
Some field theory aspects of the original M2-brane solution  with this 
Spin(7) holonomy transverse space were studied in \cite{oz}. 
In
\cite{cglp2} the fractional M2-brane  ($m\ne 0$, $\beta=0$) and a relation
to  the fractional D2-brane, which  is obtained via a reduction along the
$S^1$ isometery of the Spin(7) holonomy space, was discussed. Namely,
due to the ALC structure of the space there is now a conserved
magnetic M2-brane charge $\propto  \int_{S^4} G_4$. 
With $\beta\ne 0$ our results for the harmonic function demonstrate
that the gravitational
Chern-Simons corrections contribute to the  leading as well as the
subleading terms, along with the terms $\propto m^2$,
but with the
alternating relative signs.

\section{M2-branes with
transverse space ${\mathbf {T^*S^4}}$ }
\label{Stenzel}
Ricci-flat K\"{a}hler metrics on $T^*S^{n+1}$ were constructed for
general $n$ by
Stenzel \cite{stenzel}. Those are asymptotically conical spaces with 
the principal orbits described by a coset space $SO(n+2)/SO(n)$. 

The case of $n=2$ corresponds
to the deformed conifold with metric given originally by Candelas 
and de la Ossa \cite{candelas}. Such spaces are asymptotically
conical. 
We will specifically be interested in the case $n=3$
when the Einstein Sasakian seven manifold is $V_{5,2}=SO(5)/SO(3)$. 
The (2+1)-dimensional field theory living on the world volume of M2-branes
with transverse space $T^*S^4$ with SU(4) holonomy Stenzel metric
has $\CN=2$ supersymmetry in three dimensions. In the following, 
we find explicit M2-brane solutions with transverse space $T^*S^4$, 
taking into account the gravitational Chern-Simons $\sigma$-model
anomaly corrections. We follow closely the notation and
the explicit form of the metric for the eight-manifold as given in
\cite{cglp}.

We define left invariant 1-forms $L_{AB}$ on the group manifold
$SO(n+2)$. By splitting the index as $A=(1,2,i)$, we have that
$L_{ij}$ are the left-invariant 1-forms are the $SO(n)$ subgroup, 
and so the 1-forms in the coset $SO(n+2)/SO(n)$ will be
\ben
\sigma_i \equiv L_{1i}, ~~~~~~~\tilde{\sigma}_i \equiv L_{2i}, ~~~~~~~\nu \equiv L_{12}.
\een
The metric takes the form (for $n=4$):
\ben
\label{sten}
ds_8^2=h^2dr^2+a^2\sigma_i^2+b^2\tilde{\sigma}_i^2+c^2\nu^2,\qquad\;\; i=1,2,3.
\een
We define the Vielbeine:
\ben
e^0=h\,dr,\;\;e^i=a\,\sigma_i,\;\;e^{\tilde{i}}=b\,\tilde{\sigma}_i,\;\;e^{\tilde{0}}=c\,\nu,
\een
The functions $a$, $b$, $c$ and $h$ are given by 
\ben
\label{StSol}
& &a^2=\frac{1}{3}(2+\cosh{2r})^{1/4}\cosh{r},
\;\;\;b^2=\frac{1}{3}(2+\cosh{2r})^{1/4}\sinh{r}\,\tanh{r},\nonumber\\
& &h^2=c^2=(2+\cosh{2r})^{-3/4}\cosh^3{r}.\;
\een
As $r$ approaches zero, the metric takes the form
\[
ds^2\sim dr^2+r^2\tilde{\sigma}_i^2 + \sigma_i^2 + \nu^2
\]
which has the structure locally of the product $R^4 \times S^4$, with
$R^4$ corresponding to the "cotangent directions". As $r$ tends to 
infinity, the metric becomes
\[
ds^2 \sim d\rho^2 + \rho^2\Bigl({9 \over 16} \nu^2+{3 \over {32}}(\sigma_i^2 
+\tilde{\sigma}_i^2)\Bigr), 
\]
representing a cone over the seven-dimensional Einstein space
$V_{5,2}=SO(5)/SO(3)$.

The spin connection $\omega_{\alpha \beta}$ and the curvature two-form
$\Theta_{\alpha \beta}$ were given in \cite{cglp}. Then, using the expression
for $\Theta_{\alpha \beta}$ in  
(\ref{Seq02}), the $\sigma$-model anomaly correction to the equations 
of motion $\CX_8$ 
can be calculated (Appendix B):
\ben
& &(2\pi)^4\,{X_8}\!=\!-\frac{5}{16} 
(2385\,\cosh^{20}{r} + 10467\,\cosh^{18}{r} + 
21966\,\cosh^{16}{r} + 28296\,\cosh^{14}r
\nonumber \\
& &+ 24687\,\cosh^{12}r + 15300\,\cosh^{10}{r}
+ 6880\,\cosh^{8}r + 2216\,\cosh^{6}
{r}
\nonumber \\
& & + 486\,\cosh^{4}{r} + 65\,\cosh^{2}r + 
4)\left/[\cosh^{20}r\,(1 
+ 2\,\cosh^{2}r)^{5}]\right. .
\label{x8s}\een
In addition, we can turn on a harmonic four-form $G_4$, which was
explicitly derived  in \cite{cglp} and its magnitude is given by
 \ben
|G_{(4)}|^2=\frac{360}{\cosh^8r}.
\een
The solution to the equation of motion (\ref{KG}) for $H$ can be found
exactly and is given explicitly in Appendix B.
One of the two integration constants has been chosen to yield a non-singular
solution.
 $H$ has the following
properties:
\begin{enumerate}
\item
$r \rightarrow 0$ 
\ben
H(r)\sim c-\left(\frac{5\,m^2}{16}+\frac{145\,\beta}{24}\right)\,3^{1/4}\,r^2+
\left(\frac{35\,m^2}{96}+\frac{5365\,\beta}{432}\right)\,3^{1/4}\,r^4 +...,
\een
\item
$r \rightarrow \infty $
\begin{eqnarray}
H(r)&\sim &\left(\frac{640\,m^2}{3^7}+\frac{205\,\beta}{243}\right)\frac{1}{\rho^6}
-\frac{20480\,2^{1/3}}{28431\,3^{2/3}}\frac{m^2}{\rho^{26/3}}- \nonumber \\ && ~~~~~~~~~\left(\frac{103180 \,6^{2/3}}
{1003833}
\beta + {396800 \over 1003833}{4\over 3}^{1/3}m^2\right) \frac{1}{\rho^{34/3}}+...
\end{eqnarray}
where $\rho$ is the proper distance defined as $h\,dr\!=\!d\rho$. 
Again note that this solution describes a  the gravity dual of
the theory living on the world volume of $N$ M2-branes placed at the
conical singularity
perturbed by relevant operators,  whose conformal dimension is determined
by the subleading term.
Note that the term proportional to
$\beta$  does not contribute at this subleading order.

\end{enumerate}

\section{Conclusions}\label{Conclusions}
We have studied M2-brane solutions with special holonomy transverse
space, taking into account the gravitational Chern-Simons corrections
arising from the $\sigma$-model anomaly on the M5 brane world 
volume. For the cases when the transverse space has the (i)
original AC Spin(7) holonomy space  and (ii) Stenzel metric with SU(4)
holonomy on $T^*S^4$, we have a clear interpretation as a deformation of
the field theory on M2-branes placed at a conical singularity. Field
theory 
living on the world volume of M2-branes placed
at the tip of these cones is known for the Stenzel case \cite{field1,field2} (see
also \cite{gukov} for earlier work).
The M2-brane solution with the resolved cone as the transverse space
is perfectly smooth, and corresponds to adding a relevant operator
to the dual field theory \cite{kh}. The solution has no horizon implying
the existence of a mass gap in the field theory. 

The gravitational Chern-Simons term effectively generates a 
bulk M2-brane charge. So for the asymptotically flat cases, 
the solution still approaches $\ads{4} \times X^7$. The AdS
scale is set by the strength of the background four form turned on ($m^2$)
and by the bulk charge generated through the eight-form anomaly. 
The leading correction to $\ads{4} \times X^7$ asymptotically still
arises from the background four-form. The gravitational Chern-Simons
term contributes at higher order. Hence the interpretation of the gravity
solution in terms
of relevant operators remains as in \cite{kh}. The gravitational
Chern-Simons term effects should correspond to $1/N$
effects in the renormalization group  flow driven by addition of the
relevant
operator in the dual field theory.

\bigskip
{\noindent \bf Acknowledgments}\\

We would like to thank K. Becker, M. Becker, C. Pope and J. Poritz for discussions.
The work is supported by  the DOE grants 
and  DE-FG02-95ER40896, NATO grant 976951 (M.C.),
by the UPenn Class of 1965 Endowed Term Chair (M.C.) and Concelho Nacional
de Desenvolvimento Cient\'ifico e Tecnol\'ogico, CNPQ, Brazil (F.B.).
F.B. would like to thank Department of Physics and Astronomy, University
of Pennsylvania  for hospitality. We are grateful to the Isaac Newton
Institute for Mathematical Sciences, Cambridge  (M.C.), Albert Einstein
Institute, Potsdam (M.C.) and the Michigan
Center for Theoretical Physics, Ann Arbor (M.C. and A.N.) for support and
hospitality during the course of the work.


\renewcommand{\theequation}{\Alph{section}.\arabic{equation}}
\appendix
\section{Spin connection, curvature and Ricci tensor for Spin(7) holonomy metric}
\label{spin7}
In this appendix we calculate the spin connection, curvature two-form
and the Ricci tensor components for the metric in (\ref{newmet}) with
the veilbein given by (\ref{Sp7vp}).

\subsection*{${\mathbf{\hat{\omega}_{ab}}}$}
The spin connection $\hat{\omega}_{ab}$ satisfying $ 
d \hat{e}^a+\hat{\omega}^a{}_b \wedge \hat{e}^b=0$ is
given by:
\setcounter{equation}{0}
\begin{eqnarray}
\label{spin}
\hat{\omega}_{01}&\!=\!&-{{{a'}}\over{a\,h}}\hat{e}^1\,,\,\,\,\,\,
\hat{\omega}_{02}\!=\!-{{{a'}}\over{a\,h}} \hat{e}^2\,,\,\,\,\,\,
\hat{\omega}_{03}\!=\!-{{{b'}}\over{b\,h}} \hat{e}^3\,,\,\,\,\,\,
\hat{\omega}_{0\alpha}\!=\!-{{{c'}}\over{c\,h}}\hat{e}^\alpha,\nonumber \\
\hat{\omega}_{12}&\!=\!&{b \over {2 a^2}}\,\hat{e}^3+{{\mu^1A_{}^1+\mu^2A_{}^2} \over \sin^2\theta}-{\cot\theta
\over a} \hat{e}^2,\nonumber \\
\hat{\omega}_{13}&\!=\!&{b \over {2 a^2}}\, \hat{e}^2, \,\,\,\,\,
\hat{\omega}_{23}\!=\!-{b \over {2 a^2}}\,\hat{e}^1, \nonumber \\
\hat{\omega}_{1\alpha}&\!=\!&{a \over {2c^2}}( \sin\psi F^2_{\alpha \beta}-
\cos\psi F^1_{\alpha \beta}) \hat{e}^\beta, \nonumber\\
\hat{\omega}_{2 \alpha}&\!=\!& {a\over {2c^2}}\Bigl(-
\sin\theta F^3_{\alpha \beta}+\cos\theta \sin \psi F^1_{\alpha \beta}
+\cos\theta \cos \psi F^2_{\alpha \beta} \Bigr) \hat{e}^\beta, \nonumber \\
\hat{\omega}_{3 \alpha}&\!=\!& -{b \over {2c^2}}\, \mu^i F^i_{\alpha \beta}\,\hat{e}^\beta, 
\nonumber\\
\hat{\omega}_{\alpha \beta}&\!=\!& 
{a \over {2c^2}}\Bigl(-\sin \psi F^2_{\alpha \beta}+\cos \psi F^1_{\alpha \beta}\Bigr)
\hat{e}^1+{b \over {2c^2}} \mu^i F^i_{\alpha \beta} \hat{e}^3 \nonumber \\ & &  
\,\,\,\,+ {a \over {2c^2}}\Bigl(\sin\theta F^3_{\alpha \beta}-
\cos\theta \sin \psi F^1_{\alpha \beta} -\cos \theta \cos \psi F^2_{\alpha \beta}\Bigr)
\hat{e}^2 +
\omega_{\alpha \beta}.
\end{eqnarray}
where $\omega_{ab}$ is the spin connection on the $\sph{4}$. 
\subsection*{$\mathbf{\hat{\Theta}_{ab}}$}
The curvature two-form
$\widehat{\Theta}_{ab}=d\hat{\omega}_{ab}+\hat{\omega}_{ac}\wedge
\hat{\omega}_{cb}$ can be computed by using the spin connection calculated above:
\begin{eqnarray}
\label{forms}
\widehat{\Theta}_{01}&\!=\!&-\Bigl({{a''} \over {a\,h^2}}-\frac{a'h'}{a\,h^3}
\Bigr)\hat{e}^0\wedge\hat{e}^1
+{b\over {2a^2\,h}}\Bigl({{a'} \over a}-{{b'} \over b}\Bigr) \hat{e}^2
\wedge\hat{e}^3-\Bigl(\frac{a'}{h}-{\frac{a{c'}}{c\,h}}\Bigr)
(-\cos\psi F^1 + \sin\psi F^2), \nonumber \\
\widehat{\Theta}_{02}&\!=\!&-\Bigl({{a''} \over {a\,h^2}}-\frac{a'h'}{a\,h^3}
\Bigr)\hat{e}^0\wedge\hat{e}^2 
+{b\over {2a^2\,h}}\Bigl({{a'} \over a}-{{b'} \over b}\Bigr) \hat{e}^3
\wedge\hat{e}^1\nonumber \\&& \,\,\,\,\,\,\,\,\,+\Bigl(\frac{a'}{h}-{\frac{a{c'}}{c\,h}}\Bigr)
(-\sin\psi\,\cos\theta F^1 -\cos\psi\,\cos\theta\, F^2+\sin\theta\,F^3), \nonumber \\
\widehat{\Theta}_{03}&\!=\!& -\Bigl({{b''} \over {b\,h^2}}-\frac{b'h'}{b\,h^3}
\Bigr)\hat{e}^0\wedge\hat{e}^3+{b\over {a^2\,h}}\Bigl({{a'} \over {a}}-{{b'} \over b}\Bigr) 
\hat{e}^2\wedge\hat{e}^1+\Bigl(\frac{b'}{h}-{{b {c'}} \over {c\,h}}\Bigr)\mu^i F^i, \nonumber \\
\widehat{\Theta}_{0\alpha}&\!=\!& -\Bigl({{c''} \over {c\,h^2}}-\frac{c'h'}{c\,h^3}\Bigr) 
\hat{e}^0 \wedge \hat{e}^\alpha\nonumber\\ & & \,\,
-{1 \over {2c^2}}\Bigl({a{c'} \over {c\,h}}-{\frac{a'}{h}}\Bigr)
\Bigl(-\sin\psi \cos\theta F^1_{\alpha \beta}-\cos\psi \cos \theta F^2_{\alpha \beta}
+\sin \theta F^3_{\alpha \beta}\Bigr)\hat{e}^{2}\wedge\hat{e}^{\beta}\nonumber \\ & &\,\, 
+\Big(\frac{ac'}{2c^3\,h}-\frac{a'}{2c^2\,h}\Big)(-\cos\psi\,F_{\alpha\beta}^1
+\sin\psi\,F_{\alpha\beta}^2)\hat{e}^{1}
\wedge\hat{e}^{\beta}+{1 \over {2 c^2}} 
\Bigl(\frac{b'}{h}-{b{c'}
\over{c\,h}}\Bigr) \mu^i F^i_{\alpha \beta} \hat{e}^3 \wedge \hat{e}^\beta,
\nonumber \\
\widehat{\Theta}_{12}&\!=\!&\Bigl(
{{{b'} \over {a^2\,h}}-{{{a'}b} \over {a^3\,h}}}\Bigr)\hat{e}^0 \wedge \hat{e}^3
-\Bigl({3 \over 4} {b^2 \over a^4}-{1\over a^2}+{{a'}^2 \over a^2\,h^2}\Bigr)
\hat{e}^1 \wedge \hat{e}^2  +\Bigl(1-{a^2 \over{2c^2}} 
-{b^2 \over {2 a^2}} \Bigr)
\mu^i F^i, \nonumber \\
\widehat{\Theta}_{13}&\!=\!& \Big[{b \over 4 ac^2}\Bigl(\cos\theta \sin \psi F^1_{\alpha \beta}+\cos \theta \cos \psi F^2_{\alpha \beta}-\sin \theta F^3_{\alpha \beta} \Bigr)
\nonumber\\ & & \,\,
-{ab \over {4c^4}} \Bigl( \mu^i F^i_{\eta \alpha} (\sin \psi F^2_{\eta \beta}
-\cos \psi F^1_{\eta \beta}) \Bigr)\Big] \hat{e}^\alpha \wedge \hat{e}^\beta
\nonumber \\ & & \,\,
+\Bigl({{b'} \over {2a^2\,h}}-{{a'}b \over {2a^3\,h}}\Bigr) \hat{e}^0 \wedge \hat{e}^2
+\Bigl({b^2 \over {4a^4}}-{{a'}{b'} \over {ab\,h^2}}\Bigr) \hat{e}^1 \wedge
\hat{e}^3, \nonumber \\ 
\widehat{\Theta}_{23}&\!=\!& \Big[ {b \over {4ac^2}} (-\sin \psi F^2_{\alpha\beta} 
+\cos \psi F^1_{\alpha\beta})
\nonumber\\ & & \,\,
-{ab \over {4c^4}} \Bigl( \mu^i F^i_{\eta \beta}(
\sin\theta F^3_{\eta\alpha}-\cos\theta \sin\psi F^1_{\eta\alpha}-
\cos\theta \cos \psi F^2_{\eta\alpha} )\Bigr)\Big] \hat{e}^\alpha \wedge \hat{e}^\beta
\nonumber \\  && \,\,\,\,
-\Bigl({{b'} \over {2a^2\,h}}-{{a'}b \over {2a^3\,h}}\Bigr)\hat{e}^0 \wedge \hat{e}^1 
+\Bigl({b^2 \over {4a^4}}-{{a'}{b'} \over {ab\,h^2}}\Bigr) \hat{e}^2 \wedge
\hat{e}^3,\nonumber\\
\widehat{\Theta}_{1\alpha}&\!=\!&\Big(\frac{a'}{2c^2\,h}-\frac{ac'}{2c^3\,h}\Big)(-\cos\psi\,
F_{\alpha\beta}^1+\sin\psi\,F_{\alpha\beta}^2)\hat{e}^{0}\wedge\hat{e}^{\beta}+
\Big(\frac{a^2}{4c^4}-\frac{a'c'}{ac\,h^2}\Big)\hat{e}^{1}\wedge\hat{e}^{\alpha}
\nonumber\\ & & \,\,
+\frac{a}{2c^3}\Big(-\cos\psi\, D_\gamma F_{\alpha\beta}^1+\sin\psi\, D_{\gamma}
\,F^2_{\alpha\beta}\Big)\hat{e}^\gamma\wedge\hat{e}^\beta\nonumber\\ & & \,\,
-\Big(\frac{a^2}{4c^4}-\frac{1}{2c^2}+\frac{b^2}{4a^2c^2}\Big)\mu^i\,F^i_{\alpha\beta}\hat{e}^{2}\wedge \hat{e}^\beta
+\Big[\frac{ab}{4c^4}\mu^i\, F^i_{\alpha\eta}(-\sin\psi\,F^2_{\eta\beta}+\cos{\psi}\,F^1_{
\eta\beta})\nonumber\\ & & \,\,
-\frac{b}{4ac^2}(\sin\psi\cos\theta\,F^1_{\alpha\beta}+\cos\psi\cos\theta\,F^2_{\alpha\beta}-\sin\theta\,F^3_{\alpha\beta})\Big]\hat{e}^\beta\wedge\hat{e}^3,
\nonumber\\
\widehat{\Theta}_{2\alpha}&\!=\!&-\Big(\frac{a'}{2c^2\,h}-\frac{ac'}{2c^3\,h}\Big)
(-\sin\psi\cos\theta\,F^1_{\alpha\beta}-\cos\psi\cos\theta\,F^2_{\alpha\beta}+\sin\theta
\,F^3_{\alpha\beta})\hat{e}^0\wedge\hat{e}^\beta\nonumber\\ & & \,\,
+\Big(\frac{a^2}{4c^4}-\frac{a'c'}{ac\,h^2}\Big)\hat{e}^2\wedge\hat{e}^\alpha-
\frac{a}{2c^3}\Big(-\sin\psi\cos\theta D_\gamma F^1_{\alpha\beta}-\cos\psi\cos\theta
 D_\gamma F^2_{\alpha\beta}\nonumber\\ & & \,\,
 +\sin\theta D_\gamma F^3_{\alpha\beta}\Big)
\hat{e}^\gamma\wedge\hat{e}^\beta
+\Big(\frac{a^2}{4c^4}-\frac{1}{2c^2}+\frac{b^2}{4a^2c^2}\Big)\mu^i\,F^i_{\alpha\beta}\hat{e}^1\wedge\hat{e}^\beta\nonumber\\ & & \,\,
+\Big[\frac{ab}{4c^4}\mu^i\,F^i_{\alpha\eta}(-\sin\psi\cos\theta\,F^1_{\eta\beta}-\cos\psi\cos\theta\,F^2_{\eta\beta}+\sin\theta
\,F^3_{\eta\beta})\nonumber\\ & & \,\,
-\frac{b}{4ac^2}(-\sin\psi\,F^2_{\alpha\beta}+\cos\psi\,F^1_{\alpha\beta} )\Big]\hat{e}^\beta\wedge\hat{e}^3,
\nonumber \\ 
\widehat{\Theta}_{3\alpha}&\!=\!&-\Big(\frac{b'}{2c^2\,h}-\frac{bc'}{2c^3\,h}\Big)\mu^i
F^i_{\alpha\beta}\hat{e}^0\wedge\hat{e}^\beta+\Big[-\frac{b}{4ac^2}(
\sin\psi\cos\theta\,F^1_{\alpha\beta}\nonumber\\ & & \,\,
+\cos\psi\cos\theta\,F^2_{\alpha\beta}-\sin\theta
\,F^3_{\alpha\beta})+\frac{ab}{4c^4}\mu^i\,F^i_{\gamma\beta}(-\sin\psi\,F^2_{\gamma\alpha}
+\cos\psi\,F^1_{\gamma\alpha})\Big]\hat{e}^1\wedge\hat{e}^\beta\nonumber\\ & & \,\,
+\Big[-\frac{b}{4ac^2}(-\sin\psi\,F^2_{\alpha\beta}
+\cos\psi\,F^1_{\alpha\beta})
-\frac{ab}{4c^4}\mu^i\,F^i_{\gamma\beta}(\sin\psi\cos\theta\,F^1_{\gamma\alpha}\nonumber\\ & & \,\,
+\cos\psi\cos\theta\,F^2_{\gamma\alpha}-\sin\theta
\,F^3_{\gamma\alpha})\Big]\hat{e}^2\wedge\hat{e}^\beta+\frac{b^2}{4c^4}(\mu^i\,F^i_{\gamma
\beta})(\mu^j\,F^j_{\gamma\alpha})\hat{e}^3\wedge\hat{e}^\beta\nonumber\\ & & \,\,
-\frac{b'c'}{bc\,h^2}\hat{e}^3\wedge\hat{e}^\alpha
-\frac{b}{2c^3}\mu^i\,D_\gamma F^i_{\alpha\beta}\,
\hat{e}^\gamma\wedge\hat{e}^\beta,
\nonumber\\
\widehat{\Theta}_{\alpha\beta}&\!=\!&\Theta_{\alpha\beta}-\frac{{c'}^2}{c^2\,h^2}\,
\hat{e}^\alpha\wedge\hat{e}^\beta+\Big\{-\frac{a^2}{4c^4}(\sin\psi\,F^2_{\alpha\,\eta}-
\cos\psi\,F^1_{\alpha\,\eta})(\sin\psi\,F^2_{\beta\gamma}-\cos\psi\,F^1_{\beta\gamma})
\nonumber\\ & & \,\,
-\frac{a^2}{4c^4}(\sin\psi\cos\theta\,F^1_{\alpha\,\eta}
+\cos\psi\cos\theta\,F^2_{\alpha\,\eta}-\sin\theta\,F^3_{\alpha\,\eta})\times
\nonumber\\ & & \,\,
(\sin\psi\cos\theta\,F^1_{\beta\gamma}
+\cos\psi\cos\theta\,F^2_{\beta\gamma}-\sin\theta\,F^3_{\beta\gamma})
+\frac{b^2}{4c^4}\Big[(\mu^i\,F^i_{\alpha\,\eta})(\mu^j\,F^j_{\beta\gamma})
\nonumber\\ & & \,\,
-2(\mu^i\,F^i_{\alpha\beta})(\mu^j\,F^j_{\eta\gamma})\Big]
+\frac{a^2}{4c^4}(-\sin\psi\,F^2_{\alpha\beta}+\cos\psi\,F^1_{\alpha\beta})
(\sin\psi\,F^2_{\eta\gamma}-\cos\psi\,F^1_{\eta\gamma})\nonumber\\ & & \,\,
+\frac{a^2}{4c^4}(-\sin\psi\cos\theta\,F^1_{\alpha\beta}
-\cos\psi\cos\theta\,F^2_{\alpha\beta}+\sin\theta\,F^3_{\alpha\beta})\times
\nonumber\\ & & \,\,
(\sin\psi\cos\theta\,F^1_{\eta\gamma}
+\cos\psi\cos\theta\,F^2_{\eta\gamma}-\sin\theta\,F^3_{\eta\gamma})\Big\}\,\hat{e}^\eta\wedge\hat{e}^\gamma
\nonumber\\ & & \,\,
+\Big(\frac{a'}{c^2\,h}-\frac{ac'}{c^3\,h}\Big)
(-\sin\psi\,F^2_{\alpha\beta}+\cos\psi\,F^1_{\alpha\beta})\hat{e}^0\wedge\hat{e}^1\nonumber\\ & & \,\,
-\Big(\frac{a'}{c^2\,h}-\frac{ac'}{c^3\,h}\Big)(\sin\psi\cos\theta\,F^1_{\alpha\beta}
+\cos\psi\cos\theta\,F^2_{\alpha\beta}-\sin\theta\,F^3_{\alpha\beta})
\hat{e}^0\wedge\hat{e}^2\nonumber\\ & & \,\,
+\frac{a}{2c^3}(-\sin\psi\,D_\gamma F^2_{\alpha\beta}
+\cos\psi\,D_\gamma F^1_{\alpha\beta})\hat{e}^\gamma\wedge\hat{e}^1\nonumber\\ & & \,\,
-\frac{a}{2c^3}(\sin\psi\cos\theta\,D_\gamma F^1_{\alpha\beta}
+\cos\psi\cos\theta\,D_\gamma F^2_{\alpha\beta}-\sin\theta\,D_\gamma F^3_{\alpha\beta})
\hat{e}^\gamma\wedge\hat{e}^2\nonumber\\ & & \,\,
+\Big(\frac{b'}{c^2\,h}-\frac{bc'}{c^3\,h}\Big)\mu^i
\,F^i_{\alpha\beta}\,\hat{e}^0\wedge\hat{e}^3+\Big[\frac{b}{2ac^2}(\sin\psi\cos\theta\,F^1_{\alpha\beta}
+\cos\psi\cos\theta\,F^2_{\alpha\beta}\nonumber\\ & & \,\,
-\sin\theta\,F^3_{\alpha\beta})
+\frac{ab}{4c^4}\mu^i\,F^i_{\gamma\beta}(-\sin\psi\,F^2_{\alpha\gamma}
+\cos\psi\,F^1_{\alpha\gamma})\nonumber\\ & & \,\,
-\frac{ab}{4c^4}\mu^i\,F^i_{\alpha\gamma}(-\sin\psi\,F^2_{\gamma\beta}
+\cos\psi\,F^1_{\gamma\beta})\Big]\hat{e}^1\wedge\hat{e}^3
+\Big[\frac{b}{2ac^2}(-\sin\psi\,F^2_{\alpha\beta}
+\cos\psi\,F^1_{\alpha\beta})\nonumber\\ & & \,\,
-\frac{ab}{4c^4}\mu^i\,F^i_{\gamma\beta}(\sin\psi\cos\theta\,F^1_{\alpha\gamma}
+\cos\psi\cos\theta\,F^2_{\alpha\gamma}-\sin\theta\,F^3_{\alpha\gamma})\nonumber\\ & & \,\,
+\frac{ab}{4c^4}\mu^i\,F^i_{\alpha\gamma}(\sin\psi\cos\theta\,F^1_{\gamma\beta}
+\cos\psi\cos\theta\,F^2_{\gamma\beta}-\sin\theta\,F^3_{\gamma\beta})\Big]
\hat{e}^2\wedge\hat{e}^3\nonumber\\ & & \,\,
+\Big(\frac{1}{c^2}-\frac{a^2}{2c^4}-\frac{b^2}{2a^2c^2}\Big)\mu^i\,F^i_{\alpha\beta}\,\hat{e}^1\wedge\hat{e}^2+
\frac{b}{2c^3}\mu^i\,D_\gamma F^i_{\alpha\beta}\,\hat{e}^\gamma\wedge\hat{e}^3.
\end{eqnarray}
 $F^i\equiv\frac{1}{2}F^i_{\alpha\beta}
\,{e}^{\alpha}\wedge{e}^{\beta}$ and $D_\gamma F^i_{\alpha\beta}=\nabla_\gamma 
F^i_{\alpha\beta}+
\epsilon_{ijk}A_\gamma^jF^k_{\alpha\beta}$ is the gauge-covariant derivative of $F_{(2)}^i$,
$\nabla_\gamma$ is the Riemannian covariant derivative on $S^4$.
$\Theta_{\alpha\beta}=e^{\alpha}\wedge e^{\beta}$ is the curvature
two-form on $S^4$.

The non-zero components of the Ricci tensor in the orthonormal basis $\widehat{R}_{ab}=\widehat{R}_{acb}{}^{c}$  are
\ben
\label{Ricci}
\widehat{R}_{00}&\!=\!&-2\Big(\frac{a''}{a\,h^2}-\frac{a'h'}{a\,h^3}\Big)-\Big(\frac{b''}{b\,h^2}-
\frac{b'h'}{b\,h^3}\Big)-4\Big(\frac{c''}{c\,h^2}-\frac{c'h'}{c\,h^3}\Big),\nonumber\\ 
\widehat{R}_{11}&\!=\!&-\Big(\frac{a''}{a\,h^2}-\frac{a'h'}{a\,h^3}\Big)-\Big(\frac{3}{4}\frac{b^2}{a^4}
-\frac{1}{a^2}+\frac{{a'}^2}{a^2\,h^2}\Big)+\Big(\frac{b^2}{4a^4}-\frac{a'b'}{ab\,h^2}\Big)
+4\Big(\frac{a^2}{4c^4}-\frac{a'c'}{ac\,h^2}\Big),\nonumber\\ 
\widehat{R}_{13}&\!=\!&\frac{ab}{4c^4}\mu^i F^i_{\eta\alpha}(-\sin{\psi} 
F^2_{\eta\alpha}+\cos{\psi} F^1_{\eta\alpha}),\nonumber\\
\widehat{R}_{22}&\!=\!&\widehat{R}_{11},\nonumber\\ 
\widehat{R}_{23}&\!=\!&\frac{ab}{4c^4}\mu^i F^i_{\alpha\eta}(-\sin\psi\cos\theta\,F^1_{\eta\alpha}
-\cos\psi\cos\theta\,F^2_{\eta\alpha}+\sin\theta\,F^3_{\eta\alpha}),\nonumber\\ 
\widehat{R}_{33}&\!=\!&-\Big(\frac{b''}{b\,h^2}-\frac{b'h'}{b\,h^3}\Big)+2\Big(\frac{b^2}{4a^4}-
\frac{a'b'}{ab\,h^2}\Big)-\frac{4\,b'c'}{bc\,h^2}+\frac{b^2}{4c^4}(\mu^i F^i_{\eta\alpha})
(\mu^j F^j_{\eta\alpha}),\nonumber\\ 
\widehat{R}_{1\alpha}&\!=\!&\frac{a}{2c^3}(-\sin{\psi} 
D_\beta F^2_{\alpha\beta}+\cos{\psi} D_\beta F^1_{\alpha\beta}),\nonumber\\ 
\widehat{R}_{2\alpha}&\!=\!&\frac{a}{2c^3}(\sin\psi\cos\theta\,D_\beta F^1_{\alpha\beta}
+\cos\psi\cos\theta\,D_\beta F^2_{\alpha\beta}-\sin\theta\,D_\beta F^3_{\alpha\beta}),
\nonumber\\ 
\widehat{R}_{3\alpha}&\!=\!&-\frac{b}{2c^3}\mu^i D_\beta F^i_{\alpha\beta},\nonumber\\ 
\widehat{R}_{\alpha\beta}&\!=\!&\Big(-\frac{c''}{c\,h^2}+\frac{c'h'}{c\,h^3}+\frac{a^2}{2c^4}-
\frac{2a'c'}{ac\,h^2}-\frac{b'c'}{bc\,h^2}-\frac{3{c'}^2}{c^2\,h^2}\Big)\delta_{\alpha\beta}
+\frac{R_{\alpha\beta}}{c^2}\nonumber \\
& &-\frac{b^2}{2c^4}(\mu^iF^i_{\alpha\delta})(\mu^jF^j_{\beta\delta})
+\frac{3}{4}\frac{a^2}{c^4}(-\sin\psi F^2_{\alpha\delta}+\cos\psi F^1_{\alpha\delta})
(\sin\psi F^2_{\beta\delta}-\cos\psi F^1_{\beta\delta})\nonumber\\ 
& &+\frac{3}{4}\frac{a^2}{c^4}(-\sin\psi\cos\theta\,F^1_{\alpha\delta}
-\cos\psi\cos\theta\,F^2_{\alpha\delta}+\sin\theta\,F^3_{\alpha\delta})\times\nonumber\\
& &(\sin\psi\cos\theta\,F^1_{\beta\delta}
+\cos\psi\cos\theta\,F^2_{\beta\delta}-\sin\theta\,F^3_{\beta\delta}).
\een
Two of the  Ricci flat metrics
 with cohomogeneity one \cite{cglp,gpp},  are
\ben
h^2(r)=\Big(1-\frac{{\it l}^{10/3}}{r^{10/3}}\Big)^{-1},\;\; a^2(r)=b^2(r)=
\frac{9}{100}r^2\Big(1-\frac{{\it l}^{10/3}}{r^{10/3}}\Big),\;\; c^2(r)=
\frac{9}{20}r^2,
\een
obtained in \cite{gpp} and 
\ben
\label{new}
&&h^2(r)=\frac{(r-{\it l})^2}{(r-3{\it l})(r+{\it l})},\qquad a^2(r)=\frac{1}{4}(r-3{\it l})
(r+{\it l}), \nonumber\\
&&b^2(r)=\frac{{\it l}^2(r-3{\it l})(r+{\it l})}{(r-{\it l})^2},\qquad
c^2(r)=\frac{1}{2}(r^2-{\it l}^2),
\een
obtained recently for the ${\B}_8$ manifold \cite{cglp2}. 
Here $r$ is a radial coordinate defined as $r>l$ and $r>3\,l$, where $l>0$, for the first and second
 solutions respectively. 
Also, in our conventions, 
\ben
F^1=-(e^5\wedge e^6+e^4\wedge e^7), \,\,\, F^2=-(e^6\wedge e^4+e^5\wedge e^7),
\,\,\, F^3=-(e^4\wedge e^5+e^6\wedge e^7),\nonumber
\een
where $e^\alpha=(e^4,e^5,e^6,e^7)$ is the basis of the tangent-space 1-forms on the unit $S^4$.
\section{Detailed Calculations}
\setcounter{equation}{0}
\label{calcs}
In this appendix we provide details for  the calculation of the harmonic
function $H$ which is the solution to eq. (\ref{KG}).

\subsection*{Old Spin(7) holonomy metric}
For the metric \cite{bs,gpp}  given in (\ref{oldmet}), 
$X_8$ is given in (\ref{x8s7}).
Then, with $m=0$ in (\ref{KG}), we have
\ben
(\sqrt{g}h^{-2}H_1')'=(2\pi)^4\beta X_8\sqrt{g}=(2\pi)^4
\beta  X_8\,\frac{3^7}{4\times10^5}
\left(1-\frac{l^{10/3}}{r^{10/3}}\right)\,r^7.
\een
which can be solved:
\ben
\label{Seq5.2}
H(r)=\frac{4\times10^5}{3^7}\int{I_1(r)\left[\left(1-\frac{l^{10/3}}
{r^{10/3}}\right)^2\,r^7
\right]^{-1}dr}.
\een
where
 $I_1(r)$ is given by
\ben
\label{Seq5.20p}
I_1(r)&=&\frac{(2\pi)^4\,3^7\,\beta}{4\times10^5}
\int\,\left[\left(1-\frac{l^{10/3}}{r^{10/3}}\right)\,r^7\right]\,X_8\,dr.
\nonumber\\
&=&-\frac{\beta}{4\times10^6}\,(-8364\,l^{50/3}-33555\,l^{40/3}\,r^{10/3}+72390\,l^{20/3}\,r^{10}+73680\,r^{40/3}\,l^{10/3}\nonumber\\
& &-14140\,l^{10}\,r^{20/3}+73680\,r^{50/3})\Big/r^{50/3}+c_1.
\een
As $r \to l$, we have
\ben
\label{Seq6.111}
&&H(r)\sim -\frac{4\times10^3}{243\,l^5}\left[\frac{163691}{(4\times10^6)}\beta+c_1\right]\frac{1}{(r-l)}-\frac{32\times10^3}{729\,l^6}\left[\frac{163691}{(4\times10^6)}\beta+c_1\right]\ln(r-l)\nonumber\\
&&+\frac{437\times10^3}{6561\,l^7}\left[c_1-\frac{234467033}{1748\times10^6}\beta\right](r-l)-\frac{731500}{19683\,l^8}\left[c_1-\frac{8970520067}
{5852\times10^6}\beta\right](r-l)^2\nonumber\\
& &+...+c.
\een
The solution is regular at $r\!=\!\,l$ if we choose
the integration constant $c_1=-(163691/4\times10^6)\beta$, and it 
then tends to a finite constant. In fact
the full regular solution  is given by
\ben
\label{Seq5.21}
H_1(r)&=&\frac{\beta}{34904520}(-5758444\,l^4\,r^{32/3}+23942926\,r^{14}\,l^{2/3}-11848824\,l^{22/3}\,r^{22/3}\nonumber\\
& &-834309\,l^{14}\,r^{2/3}-5501349\,l^{32/3}\,r^4)\Big/l^{2/3}
\,r^{50/3}\,(r^{10/3}-l^{10/3})+c_2. 
\een
The asymptotic behavior,  as $r\to\infty$, is 
\ben
\label{Seq6.222}
H(r)\sim \frac{90011}{131220}\frac{\beta}{r^6}+\frac{53171}{102060}
\frac{\beta\,l^{10/3}}{r^{28/3}}+...+c_2.
\een
\subsection*{$\B_8$-metric}
The ALC Spin(7) holonomy metric on $\B_8$ \cite{cglp2,cglpm} is given in
(\ref{new}) and
$X_8$  is found in (\ref{newSp}). Eq. (\ref{KG}), with $m=0$  then
takes
the form:
\ben
(\sqrt{g}h^{-2}H')'=(2\pi)^4\beta X_8\sqrt{g}=(2\pi)^4
\beta  X_8\,(1/16)\,(r-3\,l)\,(r+l)\,(r^2-l^2)^2\,l.
\een
This equation can be solved to yield
\ben
\label{Seq5.1}
H(r)=16\int{I_1(r)\left[l(r-3\,l)^2\,(r+l)^4\right]^{-1}dr},
\een
where $I_1(r)$ is the first integration  of $H_1$ and is given by
\ben
\label{Seq5.10}
I_1(r)&=&\frac{(2\pi)^4\beta}{16}\int\,[(r-3\,l)\,(r+l)\,(r^2-l^2)^2\,l]\,
X_8\,dr\nonumber\\
&=&-\frac{\beta\,l^3\,}{98560}(1837308\,l^4\,r^5+3530010\,l^5\,r^4-2019600\,l^6\,r^3-4883604\,l^7\,r^2\nonumber\\
& &+2013318\,l^8\,r-442365\,l\,r^8+42350\,r^9+1914528\,l^2\,r^7-3768996\,l^3\,r^6\nonumber\\
& &+2565531\,l^9)\Big/(r-l)^{12}+c_1.
\een
At short distance, {\em i.e}., as $r\!\to\!3\,l$, we have
\ben
\label{Seq5.111}
&&H(r)\sim c_2-\frac{1}{16\,l^5}\left[\frac{10323\,\beta}{315392}+c_1
\right]\frac{1}{(r-3\,l)}
-\frac{1}{16\,l^6}\left[\frac{10323\,\beta}{315392}+\,c_1 \right]
\ln(r-3\,l)\nonumber\\
&+&\frac{5}{128\,l^7}\left[-\frac{63597\,\beta}{315392}+c_1\right](r-3\,l)+
\frac{5}{512\,l^8}\left[\frac{2727777}{1576960}\beta-c_1\right](r-3\,l)^2
+...
\een
We can see that this solution is regular at $r\!=\!3\,l$ for 
$c_1=-(10323/315392)\beta$ and tends to a finite constant. 
The regular solution is given by
\ben
\label{Seq5.11}
H(r)&=&-\frac{\beta}{35481600}(3453165\,r^{13}+205750648\,l^{13}-27625320\,r^{12}\,l+87480180\,r^{11}\,l^2\nonumber\\
&-&119709720\,r^{10}\,l^3-12234087\,r^9\,l^4+285056136\,r^8\,l^5-414630720\,r^7\,l^6\nonumber\\
&+&252973584\,r^6\,l^7-121143233\,r^5\,l^8+214217512\,r^4\,l^9-107560436\,r^3\,l^{10}\nonumber\\
&-&184835000\,r^2\,l^{11}-9583109\,r\,l^{12})\Big/l^5\,(r+l)^3\,(r-l)^{11}
\nonumber\\
&+&\frac{76737\,\beta}{1576960\,l^6}\ln\left(\frac{r+l}{r-l}\right)+c_2. 
\een
The asymptotic behavior, i.e., at $r\!\to\!\infty$, is
\ben
\label{Seq5.222}
H(r)\sim\frac{10323}{98560}\frac{\beta}{l\,r^5}+\frac{3441}{19712}
\frac{\beta}{r^6}+\frac{134199}{137984}\frac{l\beta}{r^7}+
\frac{34675}{19712}\frac{l^2\beta}{r^8}+...+c_2.
\een
\subsection*{
Stenzel metric}
The Stenzel metric on $T^* S^4$ is given in the main text by
eq.(\ref{sten}) and $X_8$ is given by eq. (\ref{x8s}).
We now look for solutions to (\ref{KG}), with $m=0$
Assuming that $H$ only depends on the coordinate $r$, 
we find that eq. (\ref{KG}) takes the form:
\ben
(\sqrt{g}h^{-2}H')'=(2\pi)^4\beta X_8\sqrt{g}=(2\pi)^4
\beta  X_8\,\sinh^3{2r}/216,
\een
This equation can be solved for $H$:
\ben
\label{Seq4}
H(r)=216\int{I_1(r)\left[\frac{(2+\cosh{2r})^{-3/4}\cosh^3{r}}{\sinh^3{2r}
}\right]dr},
\een
where $I_1(r)$ is the first integration of $H_1$ and is given by
\ben
I_1(r)&=&(2\pi)^4\beta\int\frac{\,\sinh^3{2r}}{216}\,X_8\,dr.\nonumber\\
  &=&-\frac{5\beta}{27648} \, (-715906\cosh^2{2r}-1660504\cosh{2r} +
72432\cosh^9{2r}
\nonumber\\
&+& 489417\cosh^8{2r} + 7660840\cosh^5{2r}  + 4778536\cosh^6{2r}
\nonumber\\
&+& 1920696\cosh^7{2r} + 7506142\cosh^4{2r}  + 3485960\cosh^3{2r}
\nonumber\\
&+& 4770\cosh^{10}{2r} - 587631)\left/[(\cosh{2r} + 2)^4  (\cosh{2r} + 1)^8 ]
\right.\nonumber\\ 
&+&c_1
\een
The solution  Eq.(\ref{Seq4}) for $H$ has a small  $r$ expansion: 
\ben
\label{Seq5}
& &H(r)\!\sim\!\Big\{-\frac{9}{2}\left(\frac{205\beta}{1024}+c_1\right)
\frac{3^{1/4}}{r^2} 
-9\left(\frac{205\beta}{1024}+c_1\right)\,3^{1/4}\,\ln{r}\nonumber\\
&+&\left[\left.\frac{93}{40}\left(\frac{205\beta}{1024}+c_1\right)-\frac{145\beta}{24}\right]\,
3^{1/4} r^2+
\left[-\frac{641}{1680}\left(\frac{205\beta}{1024}+c_1\right)+\frac{5365\beta}{432}\right]\,
3^{1/4} r^4\right.\nonumber\\
&+& ...\Big\}+c.
\een
Note that the solution at $r\!=\!0$ is regular if $c_1\!=\!-205\beta/1024$
 and tends 
to a finite
constant. We also see that this exactly agrees with the condition $I_1(0)=0$. 
Indeed, for this specific choice of $c_1$  the integration in
Eq.(\ref{Seq4}), can be 
performed exactly by redefining the coordinate $r$ as
$2+\cosh{2r}=y^4.$
In terms of the new coordinate $y$ we find 
\ben
&H(y)\!\!\!\!&=c_2-\frac{7995\sqrt{2}\,\beta}{1232}\,F(\arcsin\,(\frac{1}{y})|-1)
+\frac{\sqrt{2}\,\beta}{59136}\,(383760\,y^{44}-2839824\,y^{40}\nonumber\\
+&\!\!\!\!9224385\!\!\!\!&\,y^{36}-16750755\,y^{32}+18533340\,y^{28}-12563700\,y^{24}
+4910886\,y^{20}\nonumber\\
-&\!\!\!\!850770\!\!\!\!\!&\,y^{16}-29300\,y^{12}
+29340\,y^8-14175\,y^4
+6237)\Big/(y^4-1)^{15/2}\,y^{15},
\een
where $c_2$ is an integration constant, and $F(\phi|m)$ is the
incomplete elliptic integral of the first kind, i.e.,
\ben
\label{ell}
F(\phi|m)\equiv \int_0^\phi (1-m\,\sin^2\,\theta)^{-1/2}\,d\theta.
\een
 For large $r$ the 
solution can be 
expressed as 
\ben
\label{Seq6}
& &H(r)\sim\frac{205}{243}\frac{\beta}{\rho^6}-\frac{103180\,6^{2/3}}{1003833}
\frac{\beta}{\rho^{34/3}}
+...+c_2,
\een
where $\rho$ is the proper distance defined as $h\,dr\!=\!d\rho$. We can
see that there is no divergence at large distance and the M2-brane has a 
well-defined ADM mass.

\eject

\end{document}